\documentclass[showpacs,preprintnumbers,
superscriptaddress,amsmath]{revtex4}

\usepackage{epsfig}

\newcommand{\PRD}[3]{Phys.\ Rev.\ D\ {\bf #1},\ #2 (#3)}

\newcommand\rr{\rho}

\newcommand{\be}{\begin{equation}}
\newcommand{\ee}{\end{equation}}
\newcommand{\bea}{\begin{eqnarray}}
\newcommand{\eea}{\end{eqnarray}}
\newcommand{\ba}[1]{\begin{array}{#1}}
\newcommand{\ea}{\end{array}}

\begin{document}

\title{A light plasmon mode in the color-flavor-locking phase}

\author{Hossein Malekzadeh}
\email{malekzadeh@figss.uni-frankfurt.de}
\affiliation{Frankfurt International Graduate School for Science, 
J.W. Goethe-Universit\"at, D-60328 Frankfurt/Main, Germany}

\author{Dirk H.\ Rischke}
\email{drischke@th.physik.uni-frankfurt.de}
\affiliation{Institut f\"ur Theoretische Physik and
Frankfurt Institute for Advanced Studies, 
J.W. Goethe-Universit\"at, D-60328 Frankfurt/Main, Germany}

\date{\today}

\begin{abstract}
We calculate the spectral densities of electric and magnetic gluons
at zero temperature in color-superconducting quark matter in
the color-flavor-locking (CFL) phase.
We find a collective excitation, a plasmon, at energies smaller
than two times the gap parameter and momenta smaller than about
eight times the gap. The dispersion relation of this mode
exhibits a minimum at some nonzero value of momentum, indicating
a van Hove singularity.
\end{abstract}

\maketitle

Single-gluon exchange between two quarks
is attractive in the color-antitriplet channel.
Therefore, sufficiently cold and dense quark matter is a color
superconductor \cite{bailinlove}. When the quark-chemical
potential $\mu \gg \Lambda_{\rm QCD}$, asymptotic freedom
\cite{asympfreed} implies that the strong coupling constant $g$ at the scale
$\mu$ is much smaller than unity, $g(\mu) \ll 1$. This allows
a controlled calculation of the color-superconducting gap parameter
$\phi$ in the weak-coupling limit.

It is of interest to study the existence and the 
properties of collective gluonic excitations in a color
superconductor.
To this end, one has to compute the gluon self-energy, which allows
to determine the gluon spectral density.
At small temperatures $T \sim \phi \ll \mu$,
the dominant contribution
to the one-loop gluon self-energy
comes from a quark loop; it is $\sim g^2 \mu^2$, while
gluon (and ghost) loops contribute a term
$\sim g^2 T^2$ and are thus suppressed \cite{dhr2f}.
In this note, we focus on the gluon self-energy in
the color-flavor-locking (CFL) phase
which is characterized by an order parameter of the form
\begin{equation}
\Phi^{ij}_{fg} = \epsilon^{ijk}\, \epsilon_{fgh}\,
\Phi^k_h\;,\;\;\;\;,
\Phi^k_h \equiv \phi\, \delta^k_h\;,
\end{equation}
where $i,j,k$ are (fundamental) color indices and $f,g,h$ are
flavor indices.

\begin{figure}[ht]
\begin{center}
\includegraphics[width=8cm]{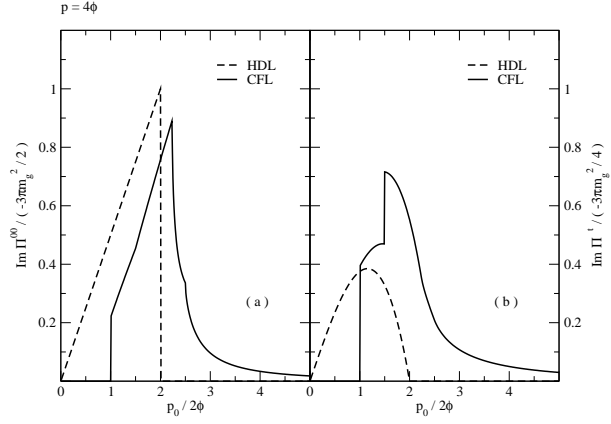}
\vspace{0.5cm}
\caption
{The imaginary part of the gluon self-energy is plotted as a function
of energy $p_{0}$ for a momentum $p=4\phi$. 
Figure (a) shows the imaginary part for electric gluons 
and (b) the corresponding one for magnetic gluons. 
The solid lines are for the CFL phase and the dotted lines for
normal-conducting matter in the HDL limit.}
\label{im00}
\end{center}
\end{figure}

To calculate the gluon self-energy in the CFL phase,
we start from Eq.\ (31a) (the self-energy of
electric gluons) and Eq.\ (31b) (the self-energy of magnetic gluons)
of Ref.\ \cite{meissner3}. We then take the zero-temperature limit
and, using the Dirac identity
\be\label{ddd}
\frac {1}{x+i\eta}={\cal P}\frac{1}{x}-i\pi\delta(x)\,\, ,
\ee
where ${\cal P}$ denotes the principal value prescription, 
we find the imaginary parts of the electric and magnetic
self-energies. The contribution from antiquarks in the one-loop
self-energies can be neglected, as the interesting range of
gluon energies is $p_{0}\ll\mu$. 
The imaginary part of the self-energies 
for electric and magnetic gluons are depicted in Fig.\
\ref{im00},
together with the results for vanishing gap parameter, which
corresponds to the hard-dense loop (HDL) limit.
As in a two-flavor color superconductor, the imaginary parts 
vanish for values of the gluon energy smaller than twice the gap
parameter $\phi$. This means that at energies smaller than $2\phi$ 
it is impossible to excite quasiparticle-quasihole pairs which 
would lead to non-vanishing imaginary parts. 
Above the light cone, for gluon energies $p_{0}> 4\phi$, 
the imaginary part of the HDL self-energy vanishes.
In the color-superconducting case, this no longer holds true,
but at least the imaginary parts decrease rapidly.

\begin{figure}[ht]
\begin{center}
\includegraphics[width=8cm]{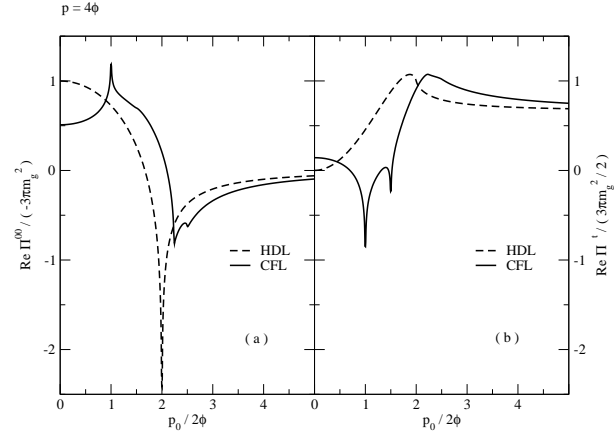}
\caption
{The real parts of the gluon self-energy as a function of energy 
$p_{0}$ for a momentum $p=4\phi$. Figure (a) shows the real part for
electric gluons and (b) the corresponding one for magnetic gluons.
The solid lines are for the 
CFL phase and the dotted lines for normal-conducting matter in
the HDL limit.}
\label{re}
\end{center}
\end{figure}

In order to compute the real parts of the gluon self-energy, one
can follow two approaches. Either, one
computes it directly from Eq.\ (\ref{ddd}) as a principal value integral,
or one employs the dispersion integral

\bea
{\rm Re}\,\Pi(p_{0},{\bf p})\equiv 
\frac{1}{\pi}{\cal P}\int_{0}^{\infty}d\omega\,
{\rm Im}\Pi(\omega,{\bf p})\Big
(\frac{1}{\omega+p_{0}}+\frac{1}{\omega-p_{0}}\Big)+C\,\, ,
\eea
where $C$ is a subtraction constant. These constants are extracted
from the behavior of the self-energies at large energies and
thus have the same values as in the normal-conducting case,
$C^{00}=0$ and $C^{t}=m_{g\hspace{.1cm}}^{2}$, for details see Ref.\
\cite{2cf}. The real parts are shown in Fig.\ \ref{re}.
\vspace{0.2cm}

Let us first consider the results for the real part of the 
electric self-energy.
In Fig.\ \ref{re}(a) there is a logarithmic singularity at 
$p_{0}=2\phi$ caused by the discontinuity of the respective 
imaginary part at the same point. 
This singularity moves toward $p_0=0$ as $\phi \to 0$.
The change of the 
gradient of the imaginary part at $p_{0}=3\phi$ produces a
corresponding change in the real part. The peak of the imaginary part
at $p_{0}=4.5\phi$, followed by a rapid decrease, produces
a cusp in the real part at the same energy. The cusp 
at $p_{0}=5\phi$ is due to the variation in the gradient of the 
imaginary part.

The discussion of the real part of the magnetic self-energy 
proceeds along similar lines. 
Because of the two discontinuities at $p_{0}=2\phi$ and $p_{0}=3\phi$ 
in Fig.\ \ref{im00}(b) there are two logarithmic singularities 
at the same energies in the real parts. Because of the change in
the gradient at $p_{0}=4\phi$ there is also a change of gradient
in the real part.
For $p_{0}\gg\phi$ the real parts of the self-energies approach 
the corresponding HDL limit. 
As one would expect, deviations from the HDL limit show up
only for $p_{0}\sim\phi$.

\begin{figure}[ht]
\begin{center}
\includegraphics[width=11cm]{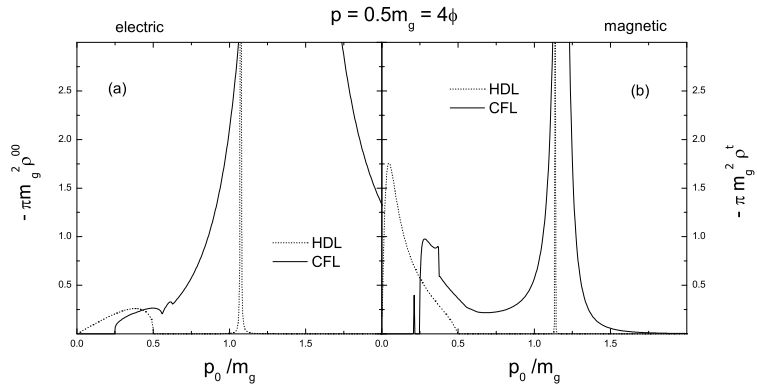}
\vspace{0.1cm}
\caption{The spectral densities for (a) electric and (b) magnetic gluons.}
\label{spd}
\end{center}
\end{figure}

From Eq.\ (46) of Ref.\ \cite{2cf} we infer the
spectral densities for the case 
${\rm Im}\Pi^{00,t}(p_{0},{\bf p})\neq 0$, 
\begin{subequations}
\bea
\rr^{00}(p_{0},{\bf p})=\frac{1}{\pi}
\frac{{\rm Im}\Pi^{00}(p_{0},{\bf p})}{
[p^{2}-{\rm Re}\Pi^{00}(p_{0},{\bf p})]^{2}
+[{\rm Im}\Pi^{00}(p_{0},{\bf p})]^{2}}\,\, ,\\
\rr^{t}(p_{0},{\bf p})=\frac{1}{\pi}
\frac{{\rm Im}\Pi^{t}(p_{0},{\bf p})}{
[p^{2}_{0}-p^{2}-{\rm Re}\Pi^{t}(p_{0},{\bf p})]^{2}
+[{\rm Im}\Pi^{t}(p_{0},{\bf p})]^{2}}\,\, .
\eea
\end{subequations}
If  ${\rm Im}\Pi^{00,t}(p_{0},{\bf p})= 0$, the spectral densities
have simple poles. For a given gluon momentum
$p$, for electric gluons the pole is determined by
\bea
[p^{2}-{\rm Re}\Pi^{00}(p_{0},{\bf p})]_{p_{0}=\omega^{00}({\bf p})}=0\,\, ,
\eea
while for the magnetic gluons it is given by
\bea
[p^{2}_{0}-p^{2}-{\rm Re}\Pi^{t}(p_{0},{\bf
p})]_{p_{0}=\omega^{t}({\bf p})}
=0\,\, .
\eea
The results are shown in Fig.\ \ref{spd}.

\begin{figure}[ht]
\begin{center}
\includegraphics[width=8cm]{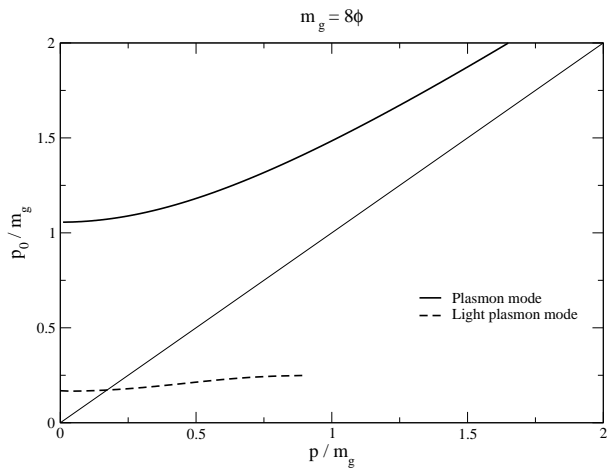}
\vspace{0.1cm}
\caption
{The plasmon dispersion relation for the regular plasmon above
the light cone and the very light plasmon mode.}
\label{plasmon1}
\end{center}
\end{figure}

We observe a peak in the magnetic spectral density at energies below
$2 \phi$, which corresponds to a collective excitation with a rather
small mass, i.e., to a very light plasmon.
The dispersion relation of this mode is shown in Fig.\ \ref{plasmon1}.
It was already predicted in Ref.\
\cite{plasmon} and exists only for energies smaller
than two times the gap parameter and momenta smaller than about 
eight times the gap. 
It exhibits a minimum at some nonzero value of momentum, 
indicating a van Hove singularity.
\vspace{0.5cm}


\begin{thebibliography}{99}

\bibitem{bailinlove}
D.\ Bailin and A.\ Love,
   {\it Phys.\ Rep.} {\bf 107}, 325 (1984).

\bibitem{asympfreed}
D.J.\ Gross and F.\ Wilczek,
   {\it Phys.\ Rev.\ Lett.} {\bf 30}, 1343 (1973);
H.D.\ Politzer,
   {\it Phys.\ Rev.\ Lett.} {\bf 30}, 1346 (1973).

\bibitem{dhr2f}
D.H.\ Rischke,
   {\it Phys.\ Rev.} D {\bf 62}, 034007 (2000).

\bibitem{meissner3}
D.H.\ Rischke, \PRD{62}{054017}{2000}.

\bibitem{2cf}
D.H.\ Rischke, \PRD{64}{094003}{2001}.

\bibitem{plasmon}
V.P.\ Gusynin and I.A.\ Shovkovy, Nucl.\ Phys.\ {\bf A700} 577-617 (2002).

\end{thebibliography}
\end{document}